\documentclass[usenames,dvipsnames, aps, pra, preprint, groupedaddress, amsfonts,
               amsmath, amssymb, showpacs, nofootinbib]{revtex4-1}
\usepackage{microtype}
\usepackage{graphicx}

\usepackage{epsfig}
\usepackage{svg}
\usepackage{siunitx}
\usepackage{epstopdf}
\usepackage{changepage}
\usepackage[utf8]{inputenc}
\usepackage[T1]{fontenc}
\usepackage{xcolor}
\usepackage{hyperref}

\usepackage{color}

\begin{document}
\raggedbottom
\title{Calculation of DC Stark Resonances for the Ammonia Molecule}

\author{Patrik Pirkola}
\email[]{patpirko@my.yorku.ca}
\author{Marko Horbatsch}
\email[]{marko@yorku.ca}
\affiliation{Department of Physics and Astronomy, York University, Toronto, Ontario M3J 1P3, Canada}
\date{\today}
\begin{abstract}
A model potential previously developed for the ammonia molecule 
is treated in a single-center partial-wave approximation in analogy with a
self-consistent field method developed by Moccia.
The latter was used in a number of collision studies.
The model potential is 
used to calculate dc Stark resonance parameters, i.e., resonance
positions and shifts within a single-center partial wave expansion,
using the exterior complex scaling method for the radial coordinate.
Three molecular valence orbitals are investigated for fields
along the three Cartesian coordinates, i.e., along the molecular
axis and in two perpendicular directions.
The work extends previous work on the planar-geometry water molecule 
for which non-monotonic shifts were observed.
We find such non-monotonic shifts for fields along the molecular
axis. For perpendicular fields we report the splitting of the
$1e$ orbitals into a fast- and a slow-ionizing orbital.
\end{abstract}
%
%

\maketitle
\section{Introduction}
Research into ionization of ammonia molecules ($\rm NH_3$) has been an ongoing
topic of interest with recently proposed new ideas about multiple ionization (or rather the
lack thereof) in the context of experimental fragmentation studies in proton-ammonia
collisions at intermediate and high energies~\cite{Wolff20}.
The fragmentation study puts (perhaps) in doubt the validity of an independent electron model
approach which was used in various forms to explain total (net) ionization cross sections,
as well as doubly differential cross sections which represent the emission properties 
of ionized electrons, i.e., their distribution over polar angle
and energy. Such studies of net ionization differential cross sections were
reported in the Born approximation~\cite{Senger1988}, and a continuum distorted wave method~\cite{Tachino_2015},
which described the earlier measurements~\cite{Lynch1976} quite well.
The fact that these experimental data were consistent with net ionization 
was demonstrated in yet another study by showing the contributions from particular
molecular orbitals (MOs)~\cite{Purkait2017}. Most of these studies employed the
single-centre Slater-type orbital based Hartree-Fock calculations of Moccia~\cite{mocc04}.

From a theoretical perspective the role of multiple ionization in proton collisions with the `isoelectronic' molecules
water ($\rm H_2O$)~\cite{Werner95,Gobet04,PhysRevA.92.032714}, methane ($\rm CH_4$)~\cite{Luna19}, and ammonia~\cite{Wolff20} 
was analyzed in the framework of the independent-atom model~\cite{PhysRevA.106.022813}.
An accurate representation of proton-water molecule differential cross sections at an intermediate energy (250 keV) was obtained with a classical trajectory Monte-Carlo method~\cite{PhysRevA.105.062822} which was based on a three-center model potential.
Collision calculations using this model potential have been carried out recently for the ammonia molecule~\cite{JHK2023}.
These works, and the problem with interpreting fragmentation cross sections~\cite{Wolff20} serve as a motivation
to extend our previous studies of Stark resonance parameters for the water molecule within a model potential approach
to the case of the ammonia molecule. The main idea of the model potential approach is to avoid the technical difficulties
of a self-consistent effective potential.

We are studying the molecule for fixed orientation, i.e., the rotational (and vibrational) degrees
of freedom are ignored. In collision problems at intermediate and high energies this approach is justified by the time scale of the collision process, and orientation averaging is applied
when computing probabilities or cross sections,
such as, e.g., in Refs.~\cite{PhysRevA.105.062822, PhysRevA.106.022813}. For the dc Stark problem
the fixed orientation with respect to the external field does represent a more serious issue, since
over longer time scales the field would act to orient a molecule with permanent dipole 
moment~\cite{UltracoldMolecules,PhysRevLett.68.1299}.
An exception would be if the molecule was found in a matrix isolation environment, e.g., by
trapping in a cold rare gas matrix. This subject has received recently renewed attention in the
context of proposals to measure the electron electric dipole moment using 
diatomic molecules~\cite{li2023baf,KOYANAGI2023111736}. Given that strong electric fields are potentially going to be applied
(cf. Ref.~\cite{PhysRevA.98.032513}) the problem of Stark ionization should be researched in this context, as well.

The Stark resonance problem is addressed in this paper by following the exterior complex scaling (ECS) method
which implements a derivative discontinuity at the radial distance where complex scaling sets in~\cite{se93}. 
The ECS methodology was developed over the years and has been compared  to the complex absorbing potential (CAP) method~\cite{Riss_1993, Santra2006}.
Following Moiseyev~\cite{Moiseyev_1998, Moiseyev_2011} one can argue that the smooth ECS and CAP methods are equivalent.
They share the features that starting at some critical radial distance $r_s$ either a gradual continuation of the real
$r$-axis into the complex plane is carried out, or a complex absorber is implemented for $r>r_s$.
Both methods show some dependence on either the scaling angle $\theta_s$ which extends the path into the complex plane,
or on the strength parameter of the CAP.  Perturbative corrections can be employed in the case of the CAP.

The hard ECS method was developed further by Scrinzi~\cite{PhysRevA.81.053845} as an effective absorber
for time-dependent problems. A derivative discontinuity in the wave function at the scaling radius $r_s$ needs to be
implemented, and it allows for the choice of scaling angles close to the critical value of $\theta_s = \pi/2$. This, in turn, allows to use a reduced
region $r_s<r<r_{\rm max}$ to  compute the tails of the resonance states.
At the outer boundary $r_{\rm max}$ the Dirichlet condition of vanishing wave function is applied.
We used his methodology previously for the planar water molecule~\cite{pm21}.
The extension from a planar geometry does not pose additional problems for the present case
of the ammonia molecule: 
a partial wave expansion of the orbitals is implemented, and as before,
the radial functions are solved using a finite element method. 
The choices for scaling radius $r_s= 16.2$ a.u., and $r_{\rm max}= 24.3$ a.u. were made in this
work combined with a scaling angle $\theta_s=0.9 \, \pi/2$.
Atomic units ($\hbar= m_e = e = 4 \pi \epsilon_0 = 1$) are used throughout this work.

\begin{figure*}[h!]
\centering

\includegraphics[width=0.5\textwidth]{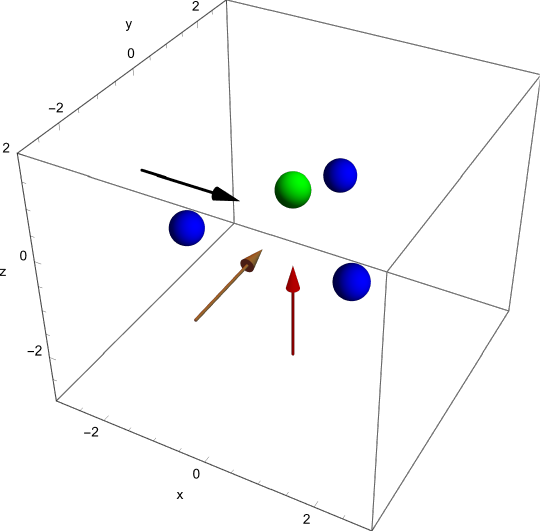}
\caption{
The geometry for the $\rm NH_3$ molecule as implemented in this work showing the nitrogen atom (in green) and
the three hydrogen atoms (in blue) schematically. Electric fields are applied pushing electrons out along the $x$-direction
(black arrow), the $y$-direction (brown arrow), and the $z$-direction (red arrow),
and will be denoted by positive values of $F_x, F_y, F_z$ respectively. Negative values of $F_x, F_y, F_z$
correspond to fields pushing in the opposite directions. The coordinates are given in atomic units.
}
\label{fig:nh3}
\end{figure*}

The geometry of the $\rm NH_3$ molecule is shown in Fig.~\ref{fig:nh3} together with the three directions along which electric fields are applied.
The arrows indicate the force directions that are applied individually, i.e., one Cartesian direction at a time. 
The force directions $F_i$ are opposite to the electric field directions $E_i$ due to the negative charge of the electron.

\section{Model}
\label{sec:model}
The model potential for the ammonia molecule is a straightforward extension of previous modelling of the water molecule
 (Refs. \cite{atoms8030059,ERREA201517,PhysRevA.83.052704,PhysRevA.99.062701,PhysRevA.102.012808}). 
 The model combines three spherically
symmetric potentials for the atomic constituents. Each part contains a screening contribution, and the 
parameters are adjusted such that the
overall potential falls of as $-1/r$ at large distances, as is required to avoid
contributions from electronic self-interaction. We keep the model parameters for the hydrogen atoms, and
model the central nitrogen atom in analogy to the oxygen atom in $\rm H_2O$.

Thus, the potential is defined as follows:
\begin{equation}\label{eq:pot0}
V_{\rm{eff}}=V_{\rm{N}}(r)+V_{\rm{H}}(r_1)+V_{\rm{H}}(r_2)+V_{\rm{H}}(r_3)\ , 
\end{equation}
\begin{align}\label{eq:pot}
\begin{split}
V_{\rm{N}}(r)=&-\frac{7-N_{\rm{N}}}{r}-\frac{N_{\rm{N}}}{r}(1+\alpha_{\rm{N}}r)\exp(-2\alpha_{\rm{N}}r)\ , \\
V_{\rm{H}}(r_j)=&-\frac{1-N_{\rm{H}}}{r_j}-\frac{N_{\rm{H}}}{r_j}(1+\alpha_{\rm{H}} r_j)\exp(-2\alpha_{\rm{H}}r_j)\ . 
\end{split}
\end{align}
The scalar variables $r_j$
(with $j=1,2,3$) represent the electron distances from the protons.
The hydrogenic parameters $\alpha_{\rm{H}}=0.6170$ and $N_{\rm{H}}=0.9075$ are taken from previous works
for the water molecule. The latter choice then fixes the potential parameter $N_{\rm{N}}=6.2775$ to yield the appropriate asymptotic effective potential at large $r$, as $3 \,(1-N_{\rm H}) = 0.2775$ is the long-range effective charge contribution to the potential from the hydrogen atoms.
The nitrogen atom screening parameter was chosen as $\alpha_{\rm{N}}=1.525$ in order for
the model to yield orbital energies that follow closely values obtained in the Hartree-Fock approximation.
The geometry of $\rm NH_3$ is adopted from the work of Moccia~\cite{mocc04}, with a N-H bond length of $1.928 \ \rm a.u.$,
polar angle $\theta_{\rm p}=108.9$ degrees, and azimuthal angles $\phi_j$ for the three hydrogenic protons spaced apart by 120 degrees.
In particular, the azimuthal angles of the hydrogen atoms are chosen to be 90, 210, 330 degrees, which singles out
the $y-z$ plane as containing one of the protons.

The Schr\"odinger equation for the MOs and an electric field in the $\hat z$ direction can be written as
\begin{equation}\label{eq:schro1}
\bigg[-\frac{1}{2}\nabla^2 - \sum_{i=0}^{3}\frac{Z_i(|\vec{r}_i|)}{|\vec{r}_i| } - F_z r  \cos(\theta)\bigg]\psi_\nu = \epsilon_\nu \,\psi_\nu,
\end{equation}
with $r_0 \equiv r$, while the $Z_i(r_i)$ are screening functions for the constituent atomic centers, i.e.,
$Z_0(r) = r V_{\rm N}(r)$ for the nitrogen atom, and $Z_i(r) = r V_{\rm H}(r)$ with $i=1,2,3$ for the hydrogens, as defined in eq.~(\ref{eq:pot}).
Note that the electric field component is $E_z=-F_z$, i.e., our notation $F_z$ refers to the force experienced by a free electron.

The MO wavefunctions $\psi \equiv \psi_\nu$ are expanded in complex-valued spherical harmonics,
\begin{equation} 
{\psi}(r, \theta, \phi) = \sum_{\ell=0}^{\ell_{\max}} \sum_{m=-\ell}^{\ell}{\sum_{i,n}^{I,N}c_{in \ell m}\frac{f_{in} (r)}{r} Y_\ell^m(\theta,\phi) } \ , 
\label{shexp}
\end{equation}
and the radial functions are expanded using a finite-element method (FEM).
The functions $f_{in}$ are local basis functions on interval $i$ of the radial interval $0 < r < r_{\max}$. 
The index $n$ labels the polynomial basis functions~\cite{se93}. The Schr\"odinger equation is solved as outlined in Ref. \cite{pm22,pm21}
and leads to a matrix eigenvalue problem with the $c_{in \ell m}$ being elements of the eigenvectors.
With this discretization technique we are solving the three-dimensional problem, which is
defined in eq.~(\ref{eq:schro1})  for a field along the ${\hat z}$ direction.
The force direction due to the external dc field as experienced by the electron is controlled by the sign of $F_z$ as explained in Fig.~\ref{fig:nh3}.

The FEM approach from Ref.~\cite{pm22,pm21}, and outlined in Ref. \cite{se93,pirkola2021exterior} was used with the partial wave expansion of $\psi$ truncated
at $\ell_{\max}=3$ to test how the MO eigenvalues respond to changes in the one free screening parameter
contained in the model potential. 
The partial wave expansion allowed the spherical components of the matrix elements to be calculated using a Wigner 3j coefficient package \cite{stone80} (which can be found on the author(s) homepage\footnote{https://www-stone.ch.cam.ac.uk/wigner.shtml}) as before in our work with water \cite{pm22,pm21,pirkola2021exterior}. 
The hydrogenic potentials are expanded in spherical harmonics which allows for the use of Wigner 3j coefficients rather than evaluating three-dimensional integrals numerically. 


Comparison with the SCF eigenvalues of Moccia shows that the three
outermost MOs can be reproduced well with the simple model potential. The $2a_1$ MO is too weakly bound
at the level of $10 \%$, which is deemed acceptable, since it is expected to contribute less to the 
overall molecular ionization rate.
The comparison of eigenvalues obtained for $\ell_{\max}=3$
and $\ell_{\max}=5$ is provided, since the resonance
parameter calculations are performed with $\ell_{\max}=3$ 
only. The table also contains results from a 
localized Hartree-Fock method as implemented in Turbomole~\cite{lhf1, lhf2, oepexx2017, tmole}.

\setlength{\tabcolsep}{6pt}
\bgroup
\def\arraystretch{0.7}
\begin{table}[!h]
\begin{center}
\begin{tabular}{c c c c c} 
\hline
\hline
   &$E_{1a1}$  & $E_{2a1}~$ &$E_{1e}$ & $E_{3a1}$ \\ [0.2ex]
   \hline
\hline
Ref.~\cite{mocc04} &-15.5222&-1.1224&-0.5956&-0.4146\\
 FEM($\ell_{\max}=3$) &-15.930&-0.976&-0.594&-0.410\\
FEM($\ell_{\max}=5$) &-15.930&-0.982&-0.609&-0.413\\
LHF &-14.087&-0.986&-0.619&-0.424\\
OEP-EXX &-14.154&-0.986&-0.611&-0.430\\
\hline
\hline
\end{tabular}
\end{center}
\caption{MO eigenvalues for the model potential as compared to the SCF eigenvalues of Moccia (Ref.~\cite{mocc04}).
The $E_{1e}$ energy appears twice, i.e. for the MOs $1e_1$ and $1e_2$.
The fourth row shows the localized HF method~\cite{lhf1,lhf2} eigenvalues based on
the optimized geometry in HF approximations as calculated in
Turbomole~\cite{tmole} using the def2-QZVPPD basis set, while
the fifth row gives the eigenvalues 
from the optimized effective potential method~\cite{oepexx2017} using the d-aug-cc-pVTZ-oep basis.}
\label{tab:intro}
\end{table}

\section{Results}

\subsection{Resonance parameters for fields along the vertical ${\pm \hat z}$-direction}

\begin{figure*}[h!]
\centering

\includegraphics[width=1.0\textwidth]{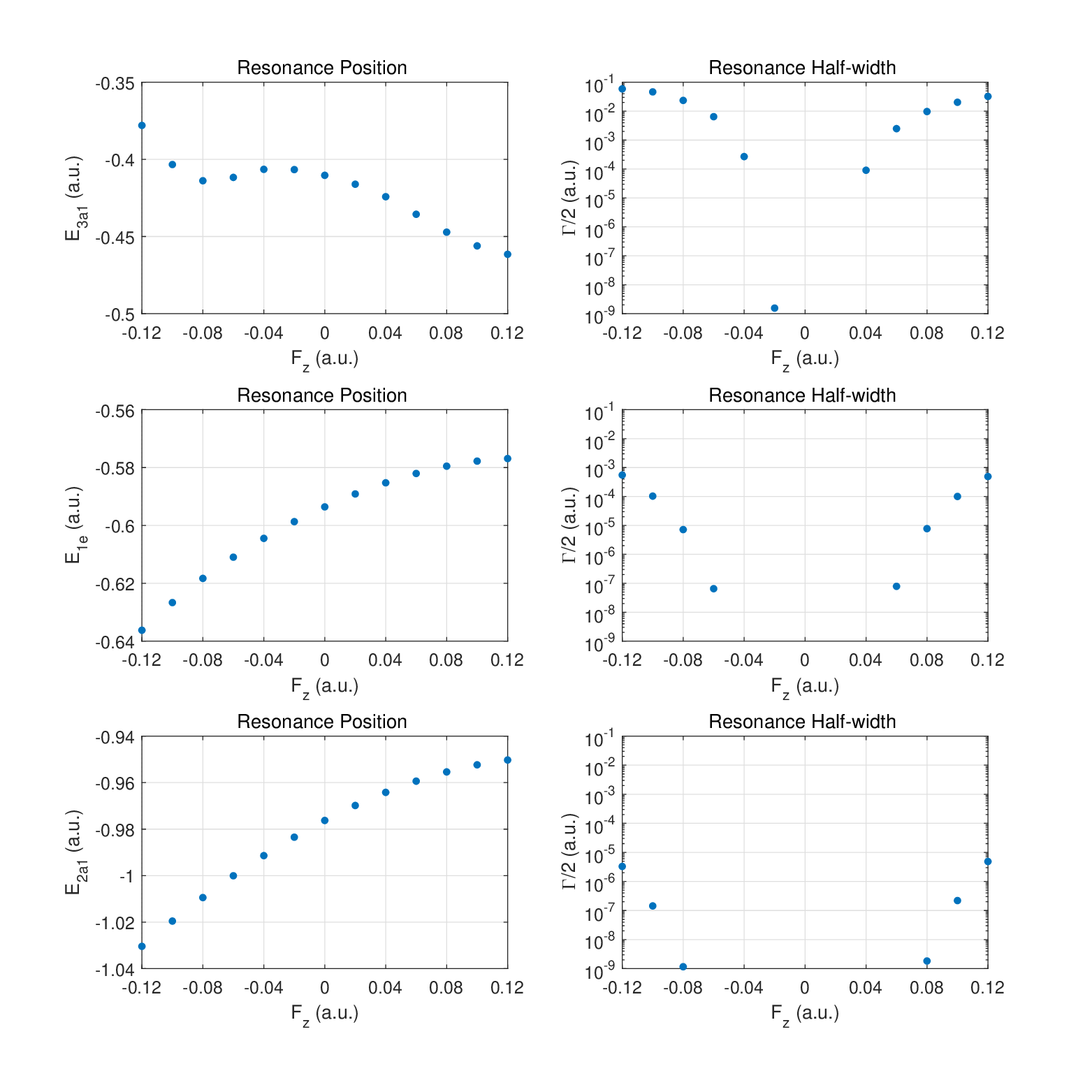}
\caption{\label{fig:zfields1}
Resonance positions (left panels) and half-widths (right panels) in atomic units for the outer
MOs for electric fields along the axis connecting the nitrogen atom with the atomic hydrogen plane.
$F_z >0$ values correspond to fields pushing electrons out on the nitrogen side,
while $F_z <0$ corresponds to emission from the side of the hydrogen atom plane.
Top row: $3a_1$, middle row: the doubly degenerate $1e$ MO (identical results for $1e_1$ and $1e_2$),
bottom row for $2a_1$.
}
\label{fig:RPalongz}
\end{figure*}

We begin the discussion of resonance parameters with fields along $\pm {\hat z}$ for which the two
degenerate MOs $1e_1$ and $1e_2$ should yield identical results. The dominant contribution to
dc field ionization is expected from the weakest bound orbital ($3a_1$).
For this orbital the combined molecular and external electric field
leads to over-the-barrier ionization at the strongest fields
calculated ($|F_z|$ of order $ 0.1 \, \rm a.u.$).

In Fig.~\ref{fig:RPalongz} the top row shows the resonance position (left column)
and resonance half-width (right column) for this orbital as a function of field strength $F_z$.
Positive values $F_z>0$ correspond to the field direction pushing electrons out in the direction
from the hydrogen atom plane past the nitrogen atom, while negative values $F_z<0$ are for electrons
ejected from the hydrogen plane (at negative $z$) away from the nitrogen atom which is located at $z=0$.

The change of the resonance position with field strength can be described as monotonically stronger binding for $F_z>0$,
since electron density is transferred from the hydrogen
atoms in the direction of the central nitrogen atom. 
For the opposite field direction ($F_z<0$)
we observe non-monotonic behavior.
First one expects marginally weaker binding when transferring 
electron density from a nitrogen to a hydrogen atom.

In the molecule the shift in electron density will be towards
regions around the partially shielded protons where the electron binding is
weaker. This feature becomes apparent at strong fields (over-the-barrier regime), but there is an intermediate range of field strengths
($-0.08 < F_z < -0.04 \,  \rm a.u.$) where there is a non-monotonic
variation of the resonance position with field strength.

The resonance widths
are obviously small in the tunneling regime. They change by orders of magnitude as the field is increased, and the ionization
rate for emission from the hydrogen plane ($F_z<0$) is stronger than in the opposite direction, by more than a
factor of two. For field strengths of the order of $0.1 \ \rm a.u.$ saturation in the ionization rate sets in, which is
associated with the over-the-barrier regime.
At these field strengths one may reach the limitations of the exponential decay model, and, thus, results for stronger fields are not reported.

We note that the behavior of the resonance position is consistent with the change in ionization rate
(or resonance width) as a function of field direction. 
In the strong field regime (at about $0.1 \ \rm a.u.$ and beyond)
the binding energies are quite different and the
ionization rates change by a factor of two when the field is reversed.
An interesting observation for $F_z>0$ is the rise in the
ionization rate even though the resonance position indicates stronger
binding. This phenomenon is associated with density being driven by
the field towards the barrier region.

In the middle panel the results are shown for the two degenerate $1e$ MOs. The dependence of the 
resonance position on field strength $F_z$ is monotonic in this case, and varies only at the $5 \%$ level
in the given field strength range. The corresponding decay rates are weaker by orders of magnitude as compared to the $3a_1$ MO,
and remain in the tunneling regime. This conclusion will be supported further below by probability density
plots for $|F_z|=0.1 \ \rm a.u.$.

The bottom panel shows results for the more deeply bound $2a_1$ MO. Here the variation in the resonance position
is only at the level of $3 \%$, and the ionization rate is suppressed by two to three additional orders of magnitude.
The shape of the dc Stark shift (left panel) as a function of field strength and orientation is similar
to what is observed for the $1e$ pair of MOs. A small asymmetry can be observed in the decay rates,
with a small enhancement for $F_z>0$ vs $F_z<0$.

In Fig.~\ref{fig:zfields2} we illustrate the situation with probability density contour plots of the MOs.
The field-free case is shown in the middle row. The outermost MO ($3a_1$) shown on the left has an
asymmetric probability density with respect to $z=0$ with higher probability values on the nitrogen side.
When the dc field is pushing electrons out on this side the nitrogen potential provides attraction,
and causes some concentration of probability in this distribution, as shown in the top left panel (strong red drop-like shape at $z>0$, and also at $z<0$). The interpretation of the density plots is that they describe steady-state decay.

The bottom left panel shows the case of a strong field pushing in the direction past the hydrogen atoms. The probability
distribution is more diffuse, showing that the outflow on the side of the hydrogen plane is hindered less.
This observation is consistent with the decay rate results shown in the top right panel of Fig.~\ref{fig:RPalongz}.
The outflow of probability density is consistent with above-the-barrier ionization for both the top and bottom rows, i.e., $|F_z|=0.1 \, \rm a.u.$.

The other two MOs show much less outflow at comparable fields,
and are clearly in the tunneling regime. 
For the $1e_1$ MO (middle column)
with field turned on in either direction there is
a limited amount of density change compared to the $3a_1$ MO. 
For the $2a_1$ MO (right column) we observe 
symmetry in the field-free case, and shifting of probability density in the direction of the applied force,
but in the tunneling regime not much probability density appears far away from the molecule.

\clearpage
\begin{figure*}[h!]
\centering

\includegraphics[width=1.0\textwidth]{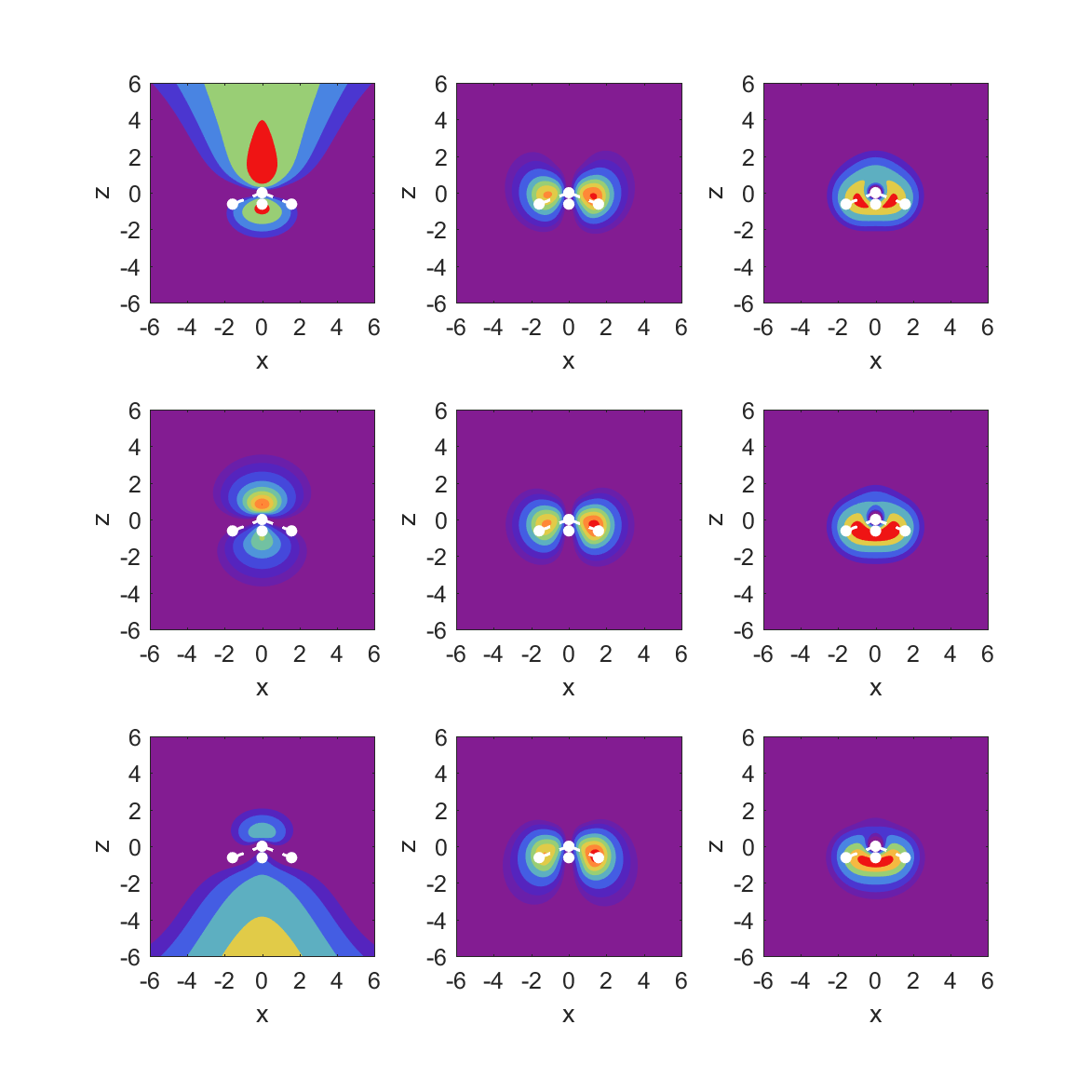}
\caption{\label{fig:zfields2}
 Probability density contour plots for the MOs
 $3a_1$, $1e_1$, $2a_1$ (left to right)
 in the $y=0$ plane, i.e., as a function of $x$ and $z$. 
 Middle row: field-free case; top row: $F_z=0.1 \ \rm a.u.$; bottom row: $F_z= -0.1 \ \rm a.u.$.
The contours are at values: $[0.00, 0.005, 0.01, 0.02, 0.04, 0.06, 0.08, 0.1, 0.12, 0.14]$.
The positions of the atomic nuclei are indicated by white dots with the N atom at $z=0$ and the
three proton locations projected onto the $x-z$ plane. The central dot corresponds to the proton residing
below the positive $y$-axis.}
\end{figure*}

\subsection{Resonance parameters for fields along 
the ${\pm \hat x}$-direction}

\begin{figure*}[h]
\centering
\includegraphics[width=1.0\textwidth]{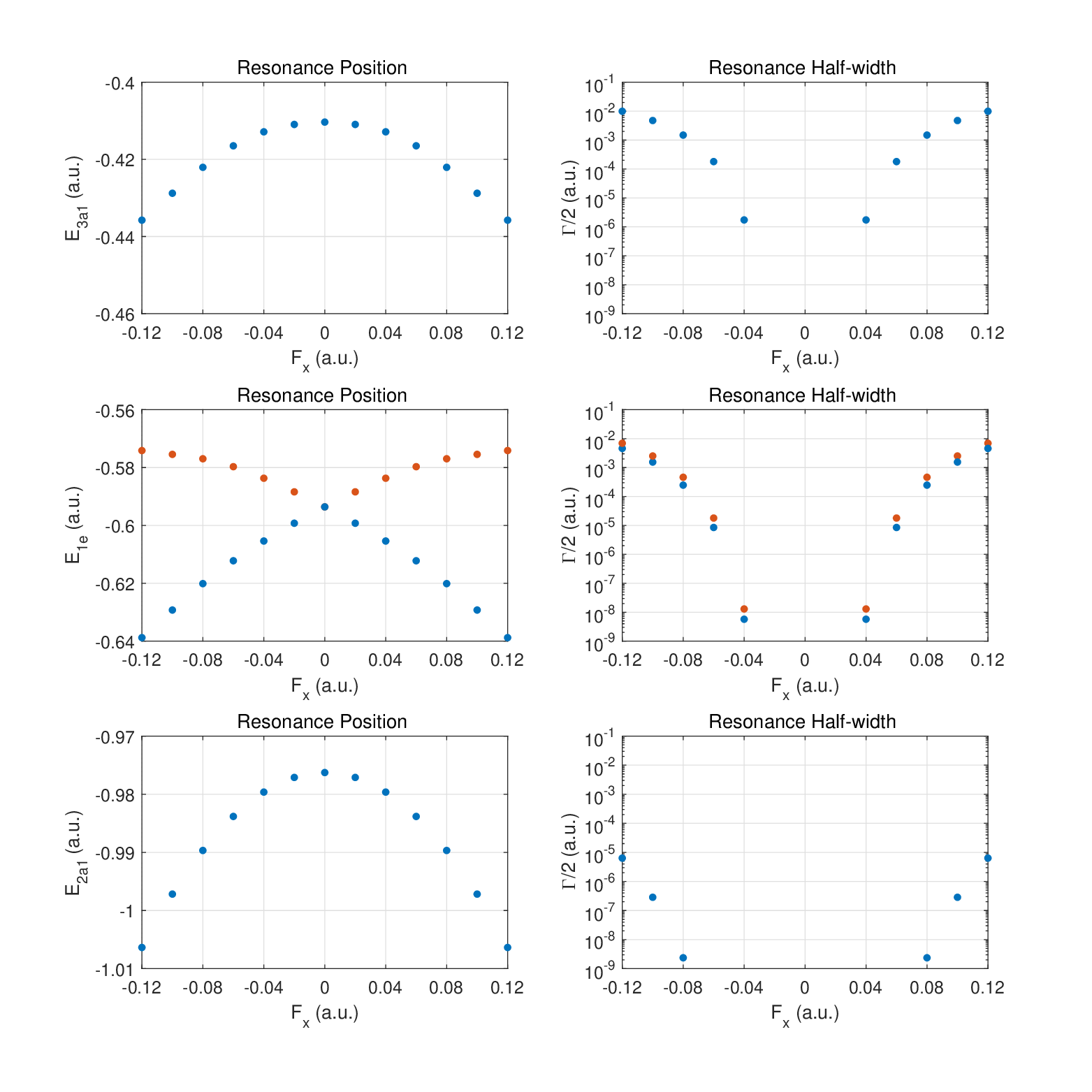}
\caption{\label{fig:RPalongx}
Same as in Fig.~\ref{fig:RPalongz} but for electric dc fields along the $x-$axis.
In the middle panel results are shown for the MOs $1e_1$ (orange dots) and $1e_2$ (blue dots).
The $1e_1$ MO has the larger ionization rate for this field orientation.}
\end{figure*}

In Fig.~\ref{fig:RPalongx} results are presented for fields in a perpendicular direction relative to the 
axis connecting the N atom with the hydrogen atom plane. 
The arrangement of the three hydrogen atoms is such
that one resides on the $y$-axis, i.e., field emission
occurs along the direction of the H-H bond perpendicular to this axis.
The degeneracy between the $1e_1$ and
$1e_2$ MO energies is expected to be broken when a dc field is applied
in this $\hat x$ (or the perpendicular $\hat y$) direction.

The top row for the outermost MO $3a_1$ shows a symmetric behaviour in the dc Stark shift (left panel)
and likewise a symmetric ionization rate with respect to 
reversal of the field direction. 
The change in the rise of the ionization rate
at strong fields indicates
that one is only approaching
the over-the-barrier regime, i.e., saturation has not set in yet
at $|F_x|=0.1 \, \rm a.u.$. 
The increase
in binding is at the $10 \%$ level for the strongest fields. The ionization rates for these fields
are smaller than for ionization along the $\hat z$ axis by a substantial factor (about three or six, depending on the 
field direction $\pm \hat z$)

The middle row shows the different behaviors for the $1e_1$ and $1e_2$ MOs.
We classify the two orbitals as fast- vs slow-ionizing under $\hat x$
oriented fields (orange vs blue markers).
The behavior is
symmetric with respect to field orientation.
The $1e_1$ MO  (orange dots) shifts towards less binding for both
field orientations, 
and the $1e_2$ (blue dots) is bound more deeply as the field is increased in either direction.

The ionization rates (right panel) are also symmetrical
with respect to field reversal.
The $1e_1$ ionizes more readily by almost a factor
of two for this field orientation. It is remarkable that these MOs ionize easily at strong fields with
the $1e_1$ MO displaying a rate which is 
moving towards that
of the more weakly bound $3a_1$ MO. Comparing the ionization rates for the $1e_1$ and $1e_2$ orbitals for fields along $\hat x$ versus
$\hat z$ we notice an order-of-magnitude increase.
The reverse trend is true for the $3a_1$ MO.

The bottom row shows that the results for the $2a_1$ MO are symmetric with respect to field orientation (as for $3a1$).
The decay rates are somewhat larger than in the case of $z$-oriented fields, even though the MO
is bound more deeply with increasing field strength.
\begin{figure*}[h!]
\centering
\includegraphics[width=1.0\textwidth]{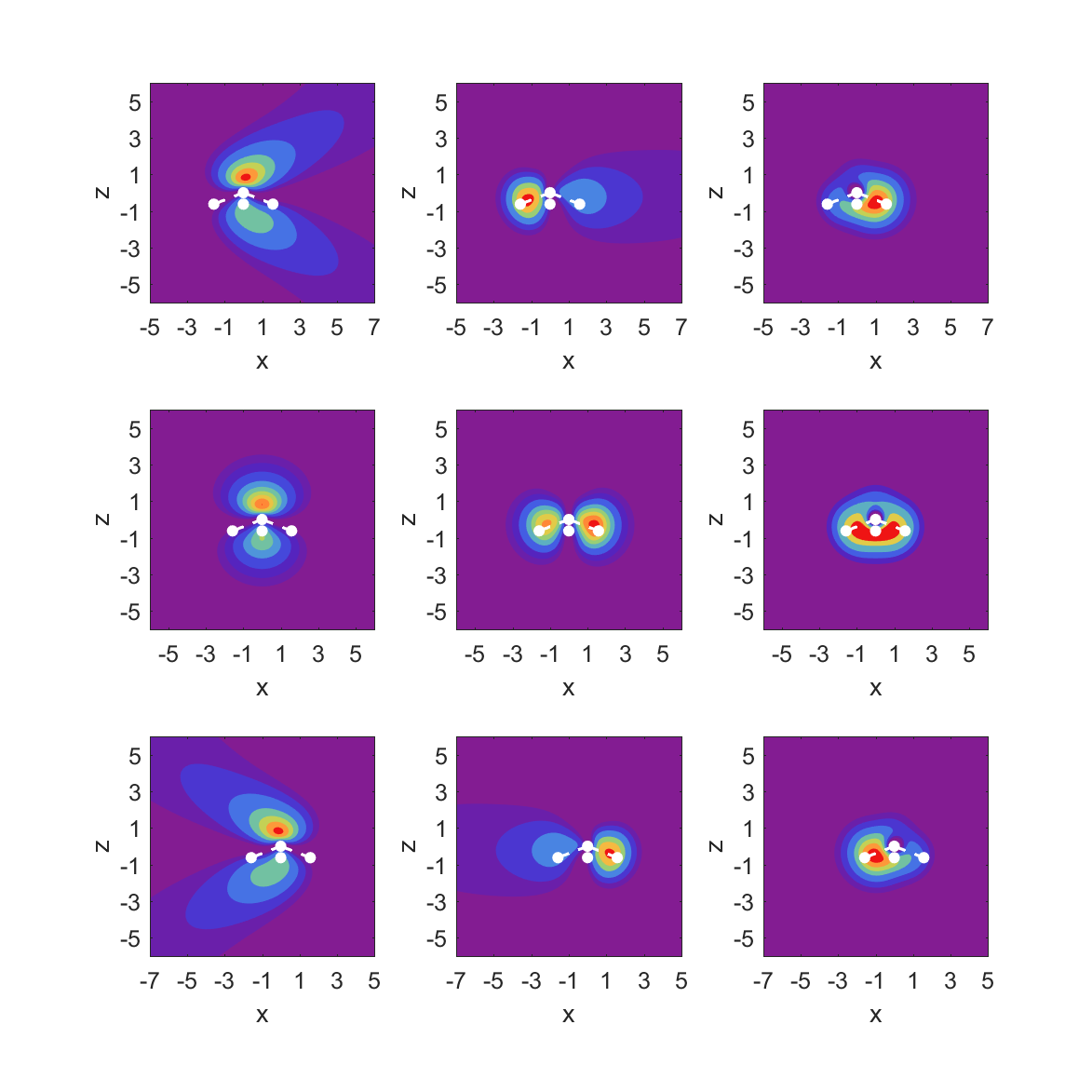}
\caption{\label{fig:xfields}
Same as in Fig.~\ref{fig:zfields2}, but for dc electric fields along the $\hat x$ axis.
Left to right: MOs $3a_1$, $1e_1$, $2a_1$; middle panel: field-free, top panel $F_x=+0.1 \ \rm a.u.$, 
bottom panel $F_x=-0.1 \ \rm a.u.$.
}
\end{figure*}

We support our resonance parameter values again with selective plots of probability densities from
the ECS approach in Fig.~\ref{fig:xfields}. The middle row is identical with that
in Fig.~\ref{fig:xfields}, but is included for direct comparison of the cases with electric field.
It is immediately apparent that three MOs ($3a_1$, $1e_1$, and also $1e_2$ which is not shown), contribute
strongly to ionization of the molecule for this field orientation.

For the $3a_1$ MO we observe outflow in the form of two jets directed above and below the hydrogen atom plane.
For the $1e_1$ MO we find that the apparent asymmetry in the $x-z$ plane for the field-free case has no
repercussions for the outflow in the case with fields of either direction: both cases show very symmetric
probabilities under these conditions which is consistent with the findings for the resonance parameters.

For the $2a_1$ MO shown in the right column symmetry with respect to field orientation is expected.
This is evident from the density plots by comparing
the top and bottom panels.
We note the relatively strong effect the field has on this relatively deeply bound orbital.

\subsection{Resonance parameters for fields along the ${\pm \hat y}$-direction}

\begin{figure*}[h!]
\centering

\includegraphics[width=1.0\textwidth]{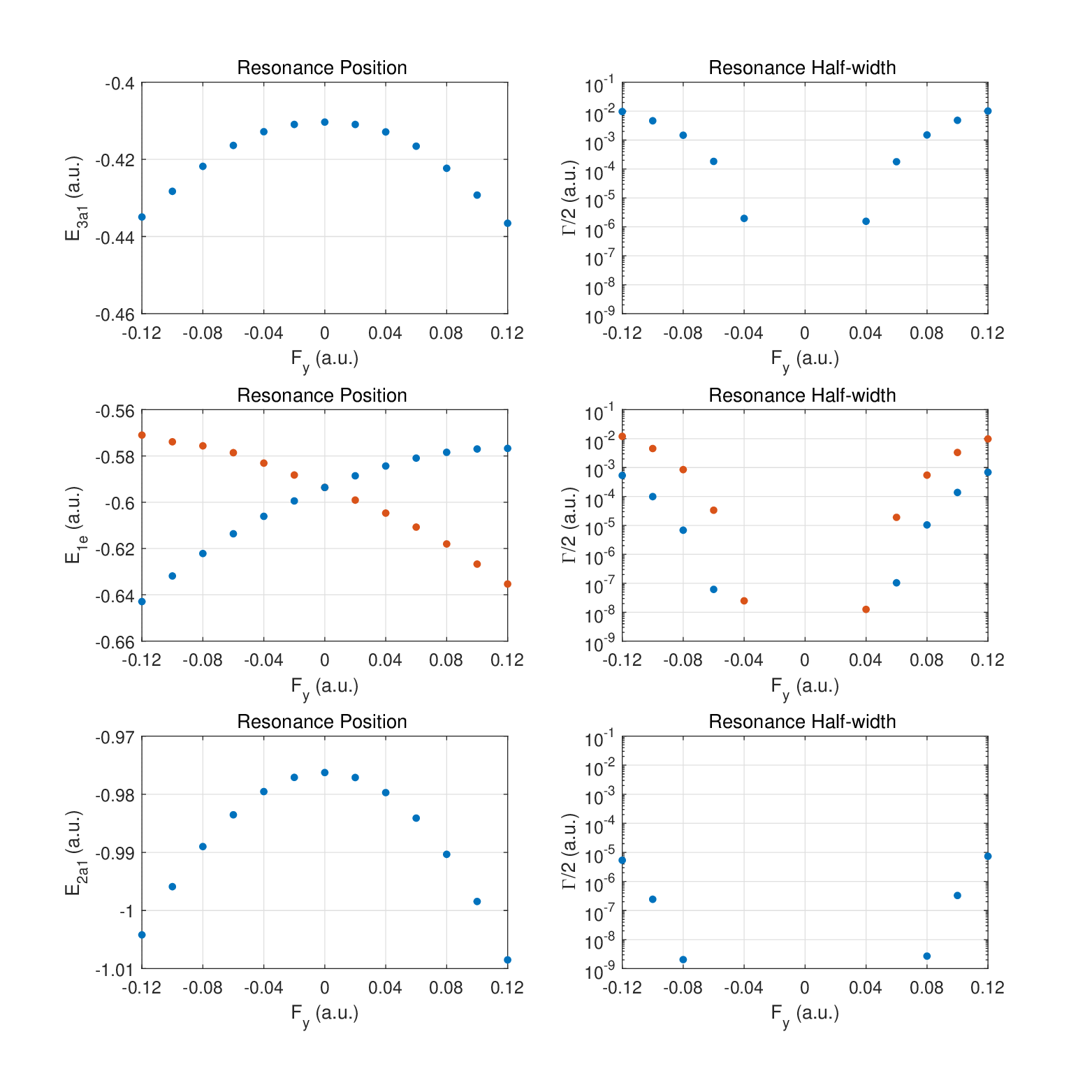}
\caption{\label{fig:RPalongy}
Same as in Fig.~\ref{fig:RPalongx} but for electric dc fields along the $y-$axis. In the middle panel results are shown for the MOs $1e_1$ (blue dots) and $1e_2$ (orange dots) which are clasified as the slow- vs fast-ionizing
$1e$ orbital respectively.
Note that the $y-z$ plane contains
a hydrogen atom at $y>0$ and this causes an asymmetry in the resonance parameters with respect to the sign of $F_y$.}

\end{figure*}

In Fig.~\ref{fig:RPalongy} results are given for field orientation along ${\hat y}$. Given the triangular nature
of the hydrogen atom plane these results differ strongly from those in the previous section. By choice of azimuthal
angles $\phi_i=90, 210, 330$ degrees, a proton is located on the positive $y$ axis, and asymmetry is expected when reversing
the field direction.

The top row shows that this asymmetry plays a very small role for the outermost MO $3a_1$ at weak fields,
and is barely noticeable.
The tabulated data (cf. Appendix C) 
show that the shifts differ by 
less than a percent.

The $1e_1$ and $1e_2$ MOs (blue and orange dots, middle row)
show markedly different behavior when compared to $\pm \hat x$
oriented fields. The shifts follow monotonic curves, as
there is no longer symmetry under field orientation reversal.
The $1e_2$ MOs is ionizing more rapidly as compared to
$1e_1$ by about two orders of magnitude. The variation
in resonance position is clearly at odds with this result,
i.e., it apparently does not play a role here.
As discussed before, the actual value of the MO binding energy
is not the deciding factor, but rather how the electron density is
driven towards the potential barrier by the external field.

The $1e_1$ MO ionizes very weakly, and its rate is
comparable to those obtained for $\pm \hat z$ oriented fields.
Thus, one may conclude that the $1e_2$ MO is affected
by this field orientation dramatically.



In the third row we give results for the deeply bound $2a_1$ MO which remains deeply in the tunneling regime
for the given field strengths. It shows a small amount of asymmetry in the dc Stark shifts and in the decay rates.

The probability density plots shown in Fig.~\ref{fig:yfields} again help to understand the finding for the parameter values. 
The results for MO $3a_1$ are very close to the corresponding plots in Fig.~\ref{fig:xfields}
and are not shown. Instead we show over the $x-y$ plane probability densities for
the strongly ionizing $1e_2$ MO (left column), the much
more weakly ionizing $1e_1$ MO
(middle column) and $2a_1$ in the right column.

For the field-free case the density plots show that the MOs $1e_2$ and $1e_1$ are actually
not perfectly aligned with our $x$- and $y$-axes. This is related to the fact that
they share the same probability density shapes, except that they are rotated by 90 degrees,
and this is incompatible with the three-fold symmetry.
Once the strong field is turned on
along either the $x$- or the $y$-axis, however, the densities respect the
symmetry of the external field, and one
can identify $1e_1 = 1e_x$ and $1e_2 = 1e_y$. 

The results for the $1e_2$ MO (left column) show an asymmetry in the outflow for the case $F_y>0$ vs $F_y<0$. 
It is interesting to observe that while the shape of the outflow 
is very different for both cases, the ionization rates are
actually differ only at the level of up to $50\, \%$
(cf. Appendix C).

\begin{figure*}[h!]
\centering
\includegraphics[width=1.0\textwidth]{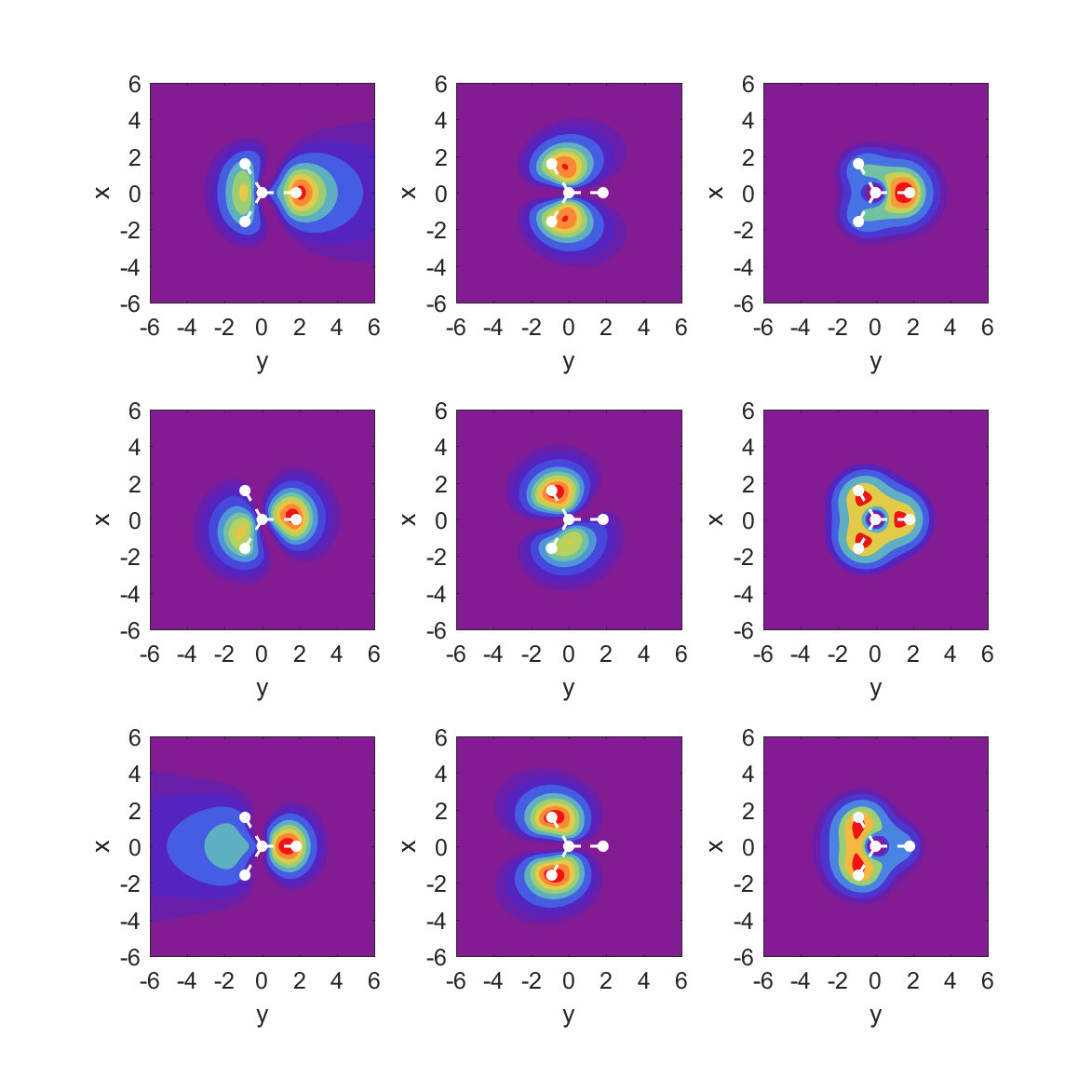}
\caption{\label{fig:yfields}
Probability densities in the $z=0$ plane for dc fields along the $\hat y$ direction.
Left column: MO $1e_2$, middle column: $1e_1$, right column: $2a_1$ are shown over the $x-y$ plane.
Middle row shows the field-free case, top row: $F_y=0.1 \ \rm a.u.$, bottom row: $F_y=-0.1 \ \rm a.u.$.
}
\end{figure*}
\newpage
In the middle column for MO $1e_1$ we observe a symmetry in the outflows despite the fact that the
arrangement of hydrogen atoms is asymmetric. For the case of MO $2a_1$ we find that the central parts of
the density are very different for the two field directions, but the parts showing the outflow are again
very similar, and this is similar to what one observes for MO $1e_1$.

\section{Conclusions}
We have extended our previous model calculation for $\rm H_2O$
to the case of $\rm NH_3$.
A simple model potential was applied to 
approximate what an SCF model (e.g., local density
functional theory)
might obtain for dc field ionization. In fact, the MO eigenvalues
were shown to be comparable to the LHF results, as shown in Table~\ref{tab:intro}.

The results demonstrate that depending
on the orientation of the electric field the three outer MOs of ammonia, i.e., $3a_1$, $1e_1$,
and $1e_2$ can have appreciable ionization rates. 
This supports the case for multiple ionization
being important whether in ion-molecule collisions or in strong-field laser-molecule interactions.

Interesting details emerge from our model calculations:
{\it (i)} non-monotonic dc shifts for the $3a1$ MO for fields along
$\hat z$; {\it (ii)} for perpendicular fields to the molecular axis
we observe that the $1e$ orbitals separate into fast- and slow-ionizing
ones; the {\it (iii)} fast-ionizing $1e$ orbital can acquire a comparable
ionization rate ti the outermost $3a1$ orbital.

It would be of interest to test these prediction with more sophisticated models, such
as Hartree-Fock theory for MO ionization rates, 
density functional theory (DFT) with electron correlation,
or even coupled-cluster theory of net ionization
which is possible within a recently developed quantum chemistry code~\cite{Jagau2014,Jagau2018,Jagau2022}.
Work is in progress to replace the model potential by
exchange-correlation potentials obtained from DFT.
\clearpage
\section{Appendix: Tables of resonance parameter values}

\subsection{Fields along $\pm \hat z$}

\setlength{\tabcolsep}{6pt}
\bgroup
\def\arraystretch{0.7}
\begin{table}[!h]
\begin{center}
\begin{tabular}{l*{2}{S[round-mode=figures,round-precision=4]}} 

 MO: $3a_{1}$ & &\\ [0.2ex] 
\hline
\hline
   \rm $\rm F_z$  & $\mathfrak{Re}$ &$\mathfrak{Im}~$ \\ [0.2ex]
   \hline
\hline

-0.12&-0.37800147483109880&-5.89258639120677635E-002\\
-0.10&-0.40335568004900579&-4.63027241968653330E-002\\
-0.08&-0.41385824214898009&-2.34901100909579649E-002\\
-0.06&-0.41169087136458904&-6.43430772683635261E-003\\
-0.04&-0.40647075602070165&-2.68648914590546384E-004\\
-0.02&-0.40663887167653390&-1.55594702628401599E-009\\
0.00&-0.41034068530840967&NA\\
0.02&-0.41610962435527843&NA\\
0.04&-0.42420845294502524&-9.04873712468544415E-005\\
0.06&-0.43558768593217162&-2.48357048163532932E-003\\
0.08&-0.44717265245611532&-9.69401778214536332E-003\\
0.10&-0.45607878840724969&-2.02729559196438078E-002\\
0.12&-0.46152511510406385&-3.22960080784600659E-002\\

\hline
\hline

\end{tabular}
\end{center}
\end{table}

\setlength{\tabcolsep}{6pt}
\bgroup
\def\arraystretch{0.7}
\begin{table}[!h]
\begin{center}
\begin{tabular}{l*{2}{S[round-mode=figures,round-precision=4]}} 

 MO: $1e$ & &\\ [0.2ex] 
\hline
\hline
   \rm $\rm F_z$  & $\mathfrak{Re}$ &$\mathfrak{Im}~$ \\ [0.2ex]
   \hline
\hline
-0.12&-0.63622429115698753&-5.52383527268262512E-004\\
-0.10&-0.62667251674709079&-1.03213832346891649E-004\\
-0.08&-0.61831181127248469&-7.17770894335906642E-006\\
-0.06&-0.61097368328799184&-6.51625820404951282E-008\\
-0.04&-0.60447199053445022&NA\\
-0.02&-0.59869609412858538&NA\\
0.00&-0.59358784772158268&NA\\
0.02&-0.58911437826738089&NA\\
0.04&-0.58526321438015105&NA\\
0.06&-0.58204485657697391&-7.85963912843962090E-008\\
0.08&-0.57951099245564719&-7.73711388518121028E-006\\
0.10&-0.57776979996196876&-1.00318500394359694E-004\\     
0.12&-0.57688023773347030&-4.91085846067488467E-004\\

\hline
\hline

\end{tabular}
\end{center}
\end{table}

\setlength{\tabcolsep}{6pt}
\bgroup
\def\arraystretch{0.7}
\begin{table}[!h]
\begin{center}
\begin{tabular}{l*{2}{S[round-mode=figures,round-precision=4]}} 

 MO: $2a_1$ & &\\ [0.2ex] 
\hline
\hline
   \rm $\rm F_z$  & $\mathfrak{Re}$ &$\mathfrak{Im}~$ \\ [0.2ex]
   \hline
\hline

-0.12&-1.0304107013885719&-3.28696225867410496E-006\\
-0.10&-1.0195482337326502&-1.43663879241951355E-007\\

-0.08&-1.0094382441693122&-1.16058817525967956E-009\\
-0.06&-1.0000563526183994&NA\\
-0.04&-0.99139635492896094&NA\\
-0.02&-0.98346320265425269&NA\\
0.00&-0.97627002352765901&NA\\

0.02&-0.96983683405716981&NA\\
0.04&-0.96419008438788301&NA\\
0.06&-0.95936288902528577&NA\\
0.08&-0.95539619467811876&-1.83416122855173507E-009\\
0.10&-0.95234205419287998&-2.19721073322533933E-007\\
0.12&-0.95027198994889095&-4.86069575089722721E-006\\

\hline
\hline

\end{tabular}
\end{center}
\end{table}

\clearpage
\subsection{Fields along $\pm \hat x$}
\vspace{6pt}
\setlength{\tabcolsep}{6pt}
\bgroup
\def\arraystretch{0.7}
\begin{table}[!h]
\begin{center}
\begin{tabular}{l*{2}{S[round-mode=figures,round-precision=4]}} 

 MO: $3a_{1}$ & &\\ [0.2ex] 
\hline
\hline
   \rm $\rm F_x$  & $\mathfrak{Re}$ &$\mathfrak{Im}~$ \\ [0.2ex]
   \hline
\hline
-0.12&-0.43576443393641351&-9.90647524258951757E-003\\
-0.10&-0.42878421012708351&-4.73992276801444169E-003\\
-0.08&-0.42206330612372206&-1.48821249140067887E-003\\
-0.06&-0.41648555454168529&-1.80270107145133159E-004\\
-0.04&-0.41285880721506624&-1.73558417105456906E-006\\
-0.02&-0.41094706580397400&NA\\
0&-0.41034068531179041&NA\\
0.02&-0.41094705264187104&NA\\
0.04&-0.41285878073028220&-1.73557794487052772E-006\\
0.06&-0.41648551433758335&-1.80270001119612141E-004\\
0.08&-0.42206325159324809&-1.48821194926094928E-003\\
0.10&-0.42878414062298387&-4.73992095773183211E-003\\
0.12&-0.43576434906228145&-9.90647092288493396E-003\\

\hline
\hline

\end{tabular}
\end{center}
\end{table}

\setlength{\tabcolsep}{6pt}
\bgroup
\def\arraystretch{0.7}
\begin{table}[!h]
\begin{center}
\begin{tabular}{l*{2}{S[round-mode=figures,round-precision=4]}} 

 MO: $1e_1$(fast) & &\\ [0.2ex] 
\hline
\hline
   \rm $\rm F_x$  & $\mathfrak{Re}$ &$\mathfrak{Im}~$ \\ [0.2ex]
   \hline
\hline
    -0.12&-0.57411809297115091&-6.92495943277970580E-003
\\
-0.10&-0.57543030009830731&-2.51500474905253695E-003\\
-0.08&-0.57695458896829821&-4.62319442736895366E-004\\
-0.06&-0.57969315528033982&-1.79892503509860458E-005\\
-0.04&-0.58368161497819937&-1.29600149169030649E-008\\
-0.02&-0.58838679123662374&NA\\
0&-0.59358784772043727&NA\\
0.02&-0.58838637094376622&NA\\
0.04&-0.58368119908473992&-1.29591151706163550E-008\\
0.06&-0.57969274887413691&-1.79891720520929910E-005\\
0.08&-0.57695418990433023&-4.62315632087699003E-004\\
0.10&-0.57542990430267182&-2.51497893002495827E-003\\

0.12&-0.57411771175025994&-6.92488195505140940E-003\\

\hline
\hline

\end{tabular}
\end{center}
\end{table}
\setlength{\tabcolsep}{6pt}
\bgroup
\def\arraystretch{0.7}
\begin{table}[!h]
\begin{center}
\begin{tabular}{l*{2}{S[round-mode=figures,round-precision=4]}} 

 MO: $1e_2$(slow) & &\\ [0.2ex] 
\hline
\hline
   \rm $\rm F_x$  & $\mathfrak{Re}$ &$\mathfrak{Im}~$ \\ [0.2ex]
   \hline
\hline

-0.12&-0.63879141999562994&-4.59191661475444844E-003\\
-0.10&-0.62924368757002458&-1.55108214344879263E-003\\
-0.08&-0.62011574038908956&-2.47540859673274277E-004\\
-0.06&-0.61221235232340432&-8.49552192423648623E-006\\
-0.04&-0.60538808153083223&-5.72510316699989643E-009\\
-0.02&-0.59923791202230681&NA\\
0&-0.59358784772043727&NA\\
0.02&-0.59923831421533125&NA\\
0.04&-0.60538845908520900&-5.72986183121232416E-009\\
0.06&-0.61221269551261703&-8.49555440248776393E-006\\
0.08&-0.62011604301180090&-2.47542598242756395E-004\\
0.10&-0.62924394596618161&-1.55109544554373228E-003\\
0.12&-0.63879162343700646&-4.59196127335634394E-003\\

\hline
\hline

\end{tabular}
\end{center}
\end{table}

\setlength{\tabcolsep}{6pt}
\bgroup
\def\arraystretch{0.7}
\begin{table}[!h]
\begin{center}
\begin{tabular}{l*{2}{S[round-mode=figures,round-precision=4]}} 

 MO: $2a_1$& &\\ [0.2ex] 
\hline
\hline
   \rm $\rm F_x$  & $\mathfrak{Re}$ &$\mathfrak{Im}~$ \\ [0.2ex]
   \hline
\hline
-0.12&-1.0063589634272361&-6.35723585248314666E-006\\
-0.10&-0.99717753037570711&-2.85510034841644849E-007\\
-0.08&-0.98966965322693223&-2.36682525877929252E-009\\
-0.06&-0.98381911516606513&NA\\
-0.04&-0.97962968072659429&NA\\
-0.02&-0.97711069509912474&NA\\
0&-0.97627002352645120&NA\\
0.02&-0.97711064292337357&NA\\
0.04&-0.97962957790850747&NA\\
0.06&-0.98381896468418728&NA\\
0.08&-0.98966945933146766&-2.36615423880420290E-009\\
0.10&-0.99717729837679525&-2.85510102891409923E-007\\
0.12&-1.0063586995040983&-6.35726407019939198E-006\\
\hline
\hline

\end{tabular}
\end{center}
\end{table}
\clearpage
\subsection{Fields along $\pm \hat y$}

\setlength{\tabcolsep}{6pt}
\bgroup
\def\arraystretch{0.7}
\begin{table}[!h]
\begin{center}
\begin{tabular}{l*{2}{S[round-mode=figures,round-precision=4]}} 

 MO: $3a_1$& &\\ [0.2ex] 
\hline
\hline
   \rm $\rm F_y$  & $\mathfrak{Re}$ &$\mathfrak{Im}~$ \\ [0.2ex]
   \hline
\hline

-0.12&-0.43493727879816213&-9.66898851763700086E-003\\
-0.10&-0.42828366104717847&-4.64414199332613062E-003\\
-0.08&-0.42180852607717068&-1.47143307570621372E-003\\
-0.06&-0.41638860365863672&-1.83260226859940207E-004\\
-0.04&-0.41283250823447809&-1.96211356839336074E-006\\
-0.02&-0.41094380938947866&NA\\
0&-0.41034068531110462&NA\\
0.02&0.41095030756154821&NA\\
0.04&-0.41288513939023186&-1.55863513173803914E-006\\
0.06&-0.41658180502473408&-1.78803761653276746E-004\\
0.08&-0.42230954677653249&-1.50646912151114021E-003\\
0.10&-0.42926290036773196&-4.82461957968768809E-003\\
0.12&-0.43655827883109610&-1.01052135810058581E-002\\
\hline
\hline

\end{tabular}
\end{center}
\end{table}

\setlength{\tabcolsep}{6pt}
\bgroup
\def\arraystretch{0.7}
\begin{table}[!h]
\begin{center}
\begin{tabular}{l*{2}{S[round-mode=figures,round-precision=4]}} 

 MO: $1e_2$(fast)& &\\ [0.2ex] 
\hline
\hline
   \rm $\rm F_y$  & $\mathfrak{Re}$ &$\mathfrak{Im}~$ \\ [0.2ex]
   \hline
\hline
-0.12&-0.57095794270216016&-1.20092032612792408E-002\\
-0.10&-0.57385815187760381&-4.55467163404213709E-003\\
-0.08&-0.57560799412934849&-8.49590680478023649E-004\\
-0.06&-0.57857113035230756&-3.35742625606113421E-005\\
-0.04&-0.58307671416313700&-2.47636744599294055E-008\\
-0.02&-0.58821937844838923&NA\\
0&-0.59358784772109219&NA\\
0.02&-0.59905608013451650&NA\\
0.04&-0.60466915689658340&-1.25435979740671886E-008\\
0.06&-0.61072235125855545&-1.90607030100124277E-005\\
0.08&-0.61799734405441809&-5.48015011397957452E-004\\
0.10&-0.62670403764295546&-3.32525981680622347E-003\\
0.12&-0.63530965702073794&-9.77220947276484682E-003\\

\hline
\hline

\end{tabular}
\end{center}
\end{table}

\setlength{\tabcolsep}{6pt}
\bgroup
\def\arraystretch{0.7}
\begin{table}[!h]
\begin{center}
\begin{tabular}{l*{2}{S[round-mode=figures,round-precision=4]}} 

 MO: $1e_1$(slow)& &\\ [0.2ex] 
\hline
\hline
   \rm $\rm F_y$  & $\mathfrak{Re}$ &$\mathfrak{Im}~$ \\ [0.2ex]
   \hline
\hline
-0.12&-0.64290036877031398&-5.35213166138552446E-004\\
-0.10&-0.63183889469936194&-9.91873055116763796E-005\\
-0.08&-0.62213164993654269&-6.83874716774646383E-006\\
-0.06&-0.61360425040960165&-6.14956510825592176E-008\\
-0.04&-0.60606771917103697&NA\\
-0.02&-0.59941451125042700&NA\\
0&-0.59358784772109219&NA\\
0.02&-0.58855942307399600&NA\\
0.04&-0.58432582526445598&NA\\
0.06&-0.58091347366591262&-1.03967321866473485E-007\\
0.08&-0.57840584964276698&-1.04330707750904628E-005\\
0.10&-0.57696903558687174&-1.37971640796002846E-004\\
0.12&-0.57670848651073126&-6.88665229486412275E-004\\

\hline
\hline

\end{tabular}
\end{center}
\end{table}

\setlength{\tabcolsep}{6pt}
\bgroup
\def\arraystretch{0.7}
\begin{table}[!h]
\begin{center}
\begin{tabular}{l*{2}{S[round-mode=figures,round-precision=4]}} 

 MO: $2a_1$& &\\ [0.2ex] 
\hline
\hline
   \rm $\rm F_y$  & $\mathfrak{Re}$ &$\mathfrak{Im}~$ \\ [0.2ex]
   \hline
\hline
-0.12&-1.0041879317142570&-5.36745761616213677E-006\\
-0.10&-0.99589626504828965&-2.43635967381625200E-007\\
-0.08&-0.98900135226698671&-2.04548668857508097E-009\\
-0.06&-0.98353258589120796&NA\\
-0.04&-0.97954366570042717&NA\\
-0.02&-0.97709980775548944&NA\\
0&-0.97627002353230541&NA\\
0.02&-0.97712152425112353&NA\\
0.04&-0.97971551357505493&NA\\
0.06&-0.98410479917594573&NA\\
0.08&-0.99033420805114569&-2.70067292964172902E-009\\
0.10&-0.99844597353231190&-3.28692054236854447E-007\\
0.12&-1.0084947926065346&-7.36156231454256975E-006\\
\hline
\hline

\end{tabular}
\end{center}
\end{table}

\clearpage
\begin{acknowledgments}
Discussions with Tom Kirchner and Michael Haslam are gratefully acknowledged. We would also like to thank Steven Chen for support with the high performance computing server used for our calculations.
Financial support from the Natural Sciences and Engineering Research Council of Canada is gratefully acknowledged. \end{acknowledgments}

\bibliography{NH3}

\end{document}